# EXPERIMENTAL INVESTIGATION OF THE FLOW DOWNSTREAM OF A DYSFUNCTIONAL BILEAFLET MECHANICAL AORTIC VALVE

Ahmed Darwish[1], Giuseppe Di Labbio[1], Wael Saleh[1], Othman Smadi[2], Lyes Kadem[1(*)]

[1] Laboratory of Cardiovascular Fluid Dynamics; Mechanical, Industrial and Aerospace Engineering; Concordia University, Montreal, QC, Canada

[2] Department of Biomedical Engineering; Hashemite University; Zarqa, Jordan

**Summary of Author contributions:**

Design of the study: OS and LK

Experimental measurements: AD and WS

Data analysis: AD, GD and LK

Drafting of the manuscript:  AD and LK

**Acknowledgements**

This work is supported by a grant from the Natural Sciences and Engineering Research Council of Canada

* Corresponding Author:

*Address:* Laboratory of Cardiovascular Fluid Dynamics (LCFD), Department of Mechanical, Industrial and Aerospace Engineering, Concordia University, 1455 de Maisonneuve Blvd. W., Montreal, Quebec, Canada, H3G 1M8
*Tel.:* (514) 848-2424 ext. 3143
*Fax:* (514) 848-3175
*Email:* lcfd@encs.concordia.ca

# ABSTRACT

Mechanical heart valve replacement is the preferred alternative in younger patients with severe symptomatic aortic valve disease. However, thrombus and pannus formations are common complications associated with bileaflet mechanical heart valves. This leads to risks of valve leaflet dysfunctions, a life-threatening event. In this experimental study, we investigate, using time-resolved planar particle image velocimetry, the flow characteristics in the ascending aorta in the presence of a dysfunctional bileaflet mechanical heart valve. Several configurations of leaflet dysfunctions are investigated and the induced flow disturbances in terms of velocity fields, viscous energy dissipation, wall shear stress and accumulation of viscous shear stresses are evaluated. We also explore the ability of a new set of parameters, solely based on the analysis of the normalized axial velocity profiles in the ascending aorta, to detect bileaflet mechanical heart valve dysfunctions and differentiate between the different configurations tested in this study. Our results show that a bileaflet mechanical heart valve dysfunction leads to a complex spectrum of flow disturbances with each flow characteristic evaluated having its own worst case scenario in terms of dysfunction configuration. We also show that the suggested approach based on the analysis of the normalized axial velocity profiles in the ascending aorta has the potential to clearly discriminate not only between normal and dysfunctional bilealfet heart valves but also between the different leaflet dysfunction configurations. This approach could be easily implemented using phase-contrast MRI to follow up patients with bileaflet mechanical heart valves.

# INTRODUCTION

The development of bileaflet mechanical aortic heart valves (BMHVs) represents a major breakthrough in the management of patients with severe symptomatic valvular disease. They are the preferred choice for younger patients because of their superior durability, compared to bioprosthetic heart valves. However, despite significant improvements in their design and implantation techniques, BMHVs still carry some risks of structural dysfunctions leading to a partial or a total obstruction of leaflet motions. BMHV dysfunction may appear at anytime between 1 hour and 20 years (median: 44.5 months) following the primary valve replacement [1]. Its reported incidence rate ranges between 0.1% to 6.0% and it represents a major life-threatening event associated with a significantly high mortality rate in emergency (28.6%) [1–5]. BMHV dysfunction is due to thrombus formation (41%), to pannus growth (38%) or to both (21%). Other cases may include: improper valve orientation, a missing leaflet or excessively long knots during surgery. An excellent review on the topic can be found in [4] and the related references [1,5–7]. Although there is a consensus that patients with BMHV dysfunction due to pannus formation should be re-operated, the optimal management of patients with BMHV dysfunction due to thrombus formation remains controversial [1,5–7]. Choosing between debriding the thrombus or instead replacing the valve is still a subject of debate. The success rate of thrombolysis therapy is high and ranges between 60-89% depending on how strongly the thrombus is adhering to the valve leaflet [1,7]. However, the rate of recurrence of thrombus, following even a successful thrombolysis therapy, remains also high (15-31%) with a mortality rate of 6-12.5% [1]. The follow-up of patients with BMHV even after a successful thrombolysis therapy remains therefore of a paramount importance [1,7]. Doppler echocardiography is typically the first technique to evaluate the performance of a BMHV. It can provide essential information regarding the hemodynamic performance of the BMHV; i.e, maximum and mean transvalvular pressure gradients, valve effective orifice area and Doppler velocity index. Studies have, however, reported a significant variability and overlap in Doppler derived parameters, the difficulty to differentiate between valve dysfunction and prosthesis-patient mismatch and between the different grades of dysfunction [8,9]. Cinefluoroscopy is also used to confirm Doppler echocardiography findings and evaluate BMHV leaflets mobility. However, cinefluoroscopy exposes patients to X-ray radiations, increasing the risks of cancers. As a consequence, using this ionizing imaging modality for the routine follow-up of patients with BMHVs is highly questionable. This justifies the need for more fundamental investigations regarding flow disturbances induced by dysfunctional BMHVs and for developing alternative modalities for patients follow-up.

From a fluid mechanics point-of-view, several studies have investigated the flow downstream of a BMHV under normal conditions [10–14]. An excellent review can be found in [10]. However, only few studies were dedicated to investigate the flow past a dysfunctional BMHV. The simulations performed by our group and others using mesh-based methods [15–17] and mesh-free methods [18] provided a good understanding of the flow structures in the ascending aorta in the presence of a dysfunctional BMHV. The above mentioned studies showed that a dysfunctional BMHV generates asymmetric flow patterns in the ascending aorta. They also confirmed that current Doppler derived parameters are not sensitive enough to detect BMHV dysfunctions. In this study, we evaluate experimentally, using time-resolved planar particle image velocimetry measurements, the flow characteristics in the ascending aorta downstream of a dysfunctional BMHV. Several configurations of leaflet dysfunctions are investigated and the induced flow disturbances in terms of velocity fields, viscous energy dissipation, wall shear stress and accumulation of viscous shear stress on particle tracers are evaluated. Furthermore, we explore the ability of a new set of parameters to detect BMHV dysfunctions and differentiate between the different severities.

## METHODS

In this study, we have used a custom-made double-activation left heart duplicator that was previously described in details in [19]. More details regarding the duplicator can be found in Appendix A and a summary of the working fluid properties and the operating conditions is reported in Table 1.

*Experimental conditions.* The BMHV used in this study is a St. Jude Regent 27A-101 with an internal diameter of 24.9 mm. The valve is assembled and placed inside the aortic model as shown in Figure 1. The orientation of leaflets relative to the sinuses is selected based on previous studies [20], where one leaflet is always facing a sinus of Valsalva. A total of six configurations are investigated including a normal operating condition, single leaflet partial and total dysfunction and two leaflets partial dysfunction. The resulting opening angles for each case are assessed by post-processing particle image velocimetry raw images. Table 2 summarizes the different configurations tested in this study and the resulting opening angles. Partial leaflet dysfunctions are induced by restricting valve leaflet(s) opening using a small nylon coated wire with a diameter of 0.7 mm. The small diameter of the wire has the advantage of not interfering with the flow downstream of the valve. Total leaflet dysfunctions are induced by applying a layer of silicone on one leaflet while in completely closed position.

*Particle image velocimetry measurements.* Time-resolved 2D particle image velocimetry (PIV) measurements are carried out using a Nd-YLF laser with a 10 mJ output energy at 1 kHz, a 527 nm wavelength and a repetition rate range between 0.2-20 kHz (Litron Laser, England). The laser sheet is positioned as shown in Figure 1. Images are captured using a Phantom V9.1 high speed camera with 1000 fps at a full resolution of 1632x1200 pixels (Vision Research, Inc., USA). Considering a PIV double-frame capturing mode and the timing between the two frames led, in our case, to a maximum of 400 velocity snapshots during 1 s. The fluid inside the heart simulator is seeded with polyamide particles (mean diameter: 50 $\mu$m, density: 1030 kg/m$^3$). A trigger was set to start the recordings at the beginning of systole. The recordings are taken after 20 cycles have elapsed to ensure that cycle-to-cycle variations are minimized. Each captured frame contains two pairs of images where the time interval between the two pairs is set at 600 $\mu$s for cases N, SLP and NSLP and at a smaller value of 400 $\mu$s for BLP, SLT and NSLT. This is done to improve cross-correlation between the captured frames and have particle displacement within the recommended range (lower than 1/4 of the interrogation zone) [21]. DaVis 7.2 software (LaVISION GmbH, Germany) is used to post-process the recorded images where it calculates the velocity vectors from the raw images by using a multiple-pass fast Fourier transform cross-correlation with an initial 32x32 pixel interrogation window and a final 16x16 pixel interrogation window with a 50% overlap. This resulted in a spatial resolution of 0.55 mm. Spurious velocity vectors are removed (cross-correlation peak ratio < 1.5) and a median filter is applied on the resulting velocity field. The uncertainty of the velocity field is less than 5% where major uncertainty contributions are evaluated using the guidelines in [21]. Also, each recording was repeated three times to ensure the repeatability of the measurements. This led to a maximum variation in $\mu$ and $\alpha$ of 6.3% and 5.8% respectively (please see below for the definition of $\mu$ and $\alpha$).

***Evaluation of flow characteristics downstream of a dysfunctional BMHV.*** In this study, the changes in flow characteristics in the ascending aorta due to the presence of a dysfunctional BMHV have been investigated by evaluating: 1) the variation in the cross-sectional normalized axial velocity profiles along the ascending aorta; 2) the temporal evolution of space-averaged viscous energy dissipation; 3) the effect of BMHV dysfunction on the aortic wall in terms of time-averaged wall shear stress and oscillatory shear index and 4) the accumulation of viscous shear stress on advected tracers in the flow stream. All the above parameters are listed in Table 3. The reader is referred to Appendix A for more details.

***A new parameter for detection and follow-up of BMHV dysfunction.*** As discussed in the introduction, the early detection and confirmation of BMHV dysfunction is of a paramount importance. Furthermore, due to the elevated rates of reoccurrence of thrombus in mechanical heart valves, studies have advocated for a close follow-up of patients even after a successful thrombolysis therapy [22,23]. So, one of the objectives of this study is to introduce velocity based non-invasive parameters having the potential to discriminate between normal and dysfunctional BMHVs but also between the different grades of dysfunctions. For this, we compute the skewness of the normalized velocity profiles, referred henceforth as: $\mu = \dfrac{\int_{-R}^{R} \frac{r}{2R} \frac{V}{V_{AVG}} dr}{\int_{-R}^{R} \frac{V}{V_{AVG}} dr}$

Considering that a *theoretical ideal* velocity profile in the aorta should be perfectly symmetric ($\mu=0$), our hypothesis is that the skewness of the velocity profile can be used as a signature of BMHV dysfunction. Furthermore, the sign of $\mu$ should indicate which leaflet is responsible for the dysfunction. We have decided to use the normalized velocity profile in order to make the suggested new parameter flow independent. The use of the skewness of velocity profile has already been reported in the literature but for flow configurations different from the one investigated in this study [24–28].

# RESULTS

***Normalized axial velocity profiles.*** Normalized axial velocity profiles are reported at the aortic root, the sinotubular junction and a section downstream from the sinotubular junction in Figure 2. This corresponds to values of y/D of 0, 1, 1.8. One can notice how a BMHV leads to a non-physiological velocity profile in the ascending aorta. For a normal BMHV, beyond the sinotubular junction (y/D=1.8), the normalized axial velocity profile approaches a flat, physiological, configuration. However, BMHV dysfunctions lead to velocity profiles that are mostly skewed even far downstream of the aortic root. One can also notice the appearance of multiple inflexion points on the normalized velocity profiles, a necessary condition for flow instability in a shear flow. We report in Appendix B additional normalized velocity profiles at twelve different sections that will be used for subsequent analyses and that can be used by others for the validation of computational fluid dynamic codes.

Figure 3 displays the temporal evolution of space-averaged viscous energy dissipation (VED) for all the cases investigated. The insert in Figure 3 shows the systolic average for each case. The results show that even partial dysfunctions lead to a significant increase in VED compared to a normal case (p<0.05). However, there is no significant difference between all the partial dysfunction cases (SLP, NSLP and BLP). The highest VED is associated with SLT case. This shows that the orientation of the dysfunctional leaflet with respect to the sinus of Valsalva has an impact on the VED in the aorta and as a consequence on left ventricle function.

Figure 4 displays the time-averaged wall shear stress (TAWSS) and the oscillatory shear index (OSI) on both aortic walls (sinus and non-sinus walls) for all the cases investigated in this study.

The TAWSS distribution for cases N, SLP, NSLP have similar patterns and values except for the elevated values near the top of the ascending aorta on the non-sinus wall for SLP and on the sinus side for NSLP. The values of the accumulation of viscous shear stress for the normal case are in good agreement with those reported by Min Yun et al. [29]. OSI distribution displays an interesting change between SLP, NSLP and the normal case, where elevated OSI value regions (~0.5) followed by a steep drop are noticeable in the ascending aorta in SLP (sinus side wall) and NSLP (non-sinus side wall). The BLP case leads to a similar TAWSS distribution on both walls with a peak value occurring a bit downstream of the sinotubular junction (~ x/D=1.44). OSI values are relatively small for both aortic walls along the aorta. SLT and NSLT cases display significantly higher TAWSS values on both aortic walls compared to other cases. OSI distribution shows however relatively low values except for localized spikes downstream of the sinotubular junction.

Figure 5 displays the distribution of the accumulation of viscous shear stresses on particle tracers released in the flow field. Over all, four different patterns can be observed: 1) the N case has most of the tracers accumulating low shear stress values; 2) the SLP and NSLP cases display a second major peak appearing in the distribution; 3) the BLP case displays a second peak and a long tail with few tracers accumulating large values of shear stresses up to 0.9 Pa.s; 4) the SLT and NSLT cases display a flattened distribution ranging up to 0.3 Pa.s. Note however that no case led to values close to the platelet activation threshold of 3.5 Pa.s after one heartbeat [30].

Figure 6 shows, for each case, the average skewness when considering all the twelve cross-sections displayed in Appendix B. The spacing between the different sections was selected specifically as 3 mm in order to reproduce the spatial resolution obtained by standard MRI machines. We also selected sections far from the aortic valve, downstream of the sinotublar junction in order to consider conditions similar to those in phase-contrast MRI where signal losses exist just downstream of a mechanical heart valve even under normal working conditions. Overall, our results show overall that there are significant differences between BMHV dysfunctions and the normal case and among the different BMHV dysfunctions ($p<0.05$). This except for the difference between N and BLP cases since both have symmetric leaflet positions and lead to almost perfectly symmetric profiles ($\mu \approx 0$). To overcome this ambiguity, we suggest introducing a second parameter ($\alpha$) also derived from the same normalized velocity profiles. The parameter ($\alpha$) represents, on a curve $V/V_{avg}$ *vs.* x/R, the ratio of the area above the average velocity to the area below the average velocity:

$$\alpha = \frac{\int \frac{V}{V_{avg}} d\frac{x}{R} \Big|_{\frac{V}{V_{avg}}>1}}{\int \frac{V}{V_{avg}} d\frac{x}{R} \Big|_{\frac{V}{V_{avg}}<1}} .$$

A flat velocity profile is expected to have a value of $\alpha=0$, while BLP case is expected to have values significantly higher than zero since the strong eccentric lateral jets have velocity magnitudes significantly higher than the average value. Introducing this second parameter allows us now to map all the cases investigated in this study on a $\mu$-$\alpha$ plan. This is displayed on Figure 7. It appears now that all severities and configurations of BMHV can be correctly discriminated ($p<0.05$ for all cases).

Another attractive feature of the parameter $\alpha$ is that it provides a good indication on how strong the shear layers are in the flow. Indeed, higher $\alpha$ values mean the existence of significant deviations from a flat velocity profile configuration and indicate the presence of elevated velocity gradients in the flow field. Interestingly enough, since velocity gradients are mostly responsible for the viscous energy dissipation, we can anticipate a good correlation between $\alpha$ values and viscous energy dissipation. This is displayed on Figure 8. Despite the limited number of data points, one can notice a good correlation with a R-value of 0.92.

# DISCUSSION

The major findings of this study are: 1) BMHV dysfunctions lead to significant changes in flow characteristics in the ascending aorta in terms of viscous energy dissipation, wall shear stress and shear stress accumulation; 2) more interestingly, there is no clear "worst case scenario" for BMHV dysfunction configurations. Each flow characteristic investigated in this study has its own worst dysfunction configuration; 3) we introduced a simple original approach based on mapping BMHV configurations on a μ-α map. This can easily be performed by phase-contrast MRI and can represent an ideal non-invasive and radiation-free approach for the confirmation of BMHV dysfunctions and for the routine follow-up of patients after thrombolysis therapy.

***Low incidence but high complexity.*** Mechanical prosthetic heart valves have experienced decades of improvements in their design. It is expected that a mechanical prosthetic heart valve will minimally disturb the flow in the ascending aorta. The bileaflet design of modern mechanical prosthetic heart valves achieves this quite very well [10,31,32]. However, they are not free of dysfunctions that can alter their optimal performance. Under such conditions valve leaflets represent a major obstacle in the flow stream and the valve displays a complex configuration mixing the adverse effects of a severe aortic stenosis and a bicuspid valve. In this experimental study, we have explored and quantified the changes in flow characteristics in the ascending aorta due to the presence of different configurations of BMHV dysfunctions. Although it was anticipated that a BMHV dysfunction will significantly alter the flow in the aorta, an important fundamental question remained unexplored in terms of what is the worst case scenario for a BMHV dysfunction. Surprisingly, our results show that BMHV dysfunction displays a complex spectrum of effects on the flow in the ascending aorta with: 1) totally blocked leaflet configurations (SLT and NSLT), owning their significant reduction in valve orifice area, result in the highest viscous energy dissipation. This dissipated energy is unrecoverable causing an increase in left ventricle load and left ventricle myocardium stress [33]; 2) single partially blocked leaflet configurations (SLP and NSLP) represent the worst case scenario for time-averaged wall shear stress and oscillatory wall shear stress. It is well known that low TAWSS with high OSI are associated with elevated risks of atherosclerosis [34]. Following this, our results show that a large portion of the ascending aorta is exposed to risks of atherosclerosis in the presence of SLP and NSLP configurations. In comparison, the other dysfunctional cases (BLP, SLT and NSLT) mostly lead to elevated TAWSS values but with low OSI; 3) both valve leaflets partially blocked configuration (BLP) represents the worst case scenario for the accumulation of viscous shear stress and risks of platelets activation. Viscous shear stress is the major stress applying mechanical load on platelets [35]. For dysfunctional cases, AVSS values did not reach the reported threshold for platelet activation of 3.5 Pa.s [30]. However, when both leaflets are restricted in motion, about 4.71% of tracers experience elevated AVSS values ranging between 0.3 Pa.s and 0.9 Pa.s. Furthermore, for BLP case, 62.78% of tracers remained in the ascending aorta at the end of the advection period. This may expose them to additional shear stress during the following heart beats. The reported results in this study are in good agreement with previous studies in the literature. For a normally functioning BMHV, numerical simulations mostly reported the velocity fields in the ascending aorta, TAWSS and OSI on valve leaflets and AVSS [12,35–40]. Their findings are in a good agreement with our experimental results. In the case of BMHV dysfunction, only few studies [16–18] are reported in the literature and their results in terms velocity profiles and AVSS are in a good qualitative agreement with our experimental findings. A thorough quantitative comparison is still difficult because the reported numerical simulations did not include fluid-structure interaction and considered stationary BMHV leaflets.

***From Catheterization to Cinefluoroscopy to Phase-contrast MRI.*** Despite its low incidence rate, BMHV dysfunction represents a life-threatening event. As a consequence, patients with BMHV require a routine follow-up in order to evaluate valve hemodynamics and leaflets mobility. Despite some interesting attempts to use high-fidelity phonocardiography and Morlet wavelet in order to detect BMHV dysfunction [41], Doppler ultrasound remains the recommended frontline approach to evaluate the performance of BMHVs [15,42]. If a BMHV dysfunction is suspected, cinefluoroscopy is recommended to confirm leaflet suboptimal mobility. However, considering its ionizing nature and the elevated rate of recurrence of thrombus (15-31%), even following a successful thrombolysis therapy, cinefluoroscopy is not recommended as a routine technique to follow-up patients with BMHVs. As a consequence, there is a need to explore alternative non-invasive and non-ionizing techniques allowing for the routine evaluation of BMHV dysfunction. In this study, we have demonstrated that BMHV dysfunction significantly alters the flow field in the ascending aorta and that every valve configuration has a distinct signature on a $\mu$-$\alpha$ plan. Indeed, our results show that although the skewness ($\mu$) is sufficient to evaluate and distinguish between the different scenarios of single leaflet dysfunction, the case with both leaflet partial dysfunction remained challenging because of a possible overlap with a normal case. This required the addition of the $\alpha$ ratio for a clear distinction between all cases. Table 4 summarizes the expected signatures of different BMHV configurations on the $\mu$-$\alpha$ plan. This proposed approach might represent an attractive radiation-free alternative to cinefluoroscopy in order to confirm BMHV dysfunction following a Doppler echocardiography assessment. In clinical practice, the normalized velocity profiles can be easily obtained using phase-contrast MRI (PC-MRI) like in [43]. The encoding velocity ($V_{enc}$) has to be adjusted in order to avoid aliasing and an anti-aliasing correction can also be used like in [44]. After obtaining the velocity field in the ascending aorta, it is straightforward to extract velocity profiles like in [45], normalize them and determine the values of $\mu$ and $\alpha$. It is expected that our study should contribute towards promoting the use of a radiation-free multi-parametric approach (Doppler echocardiography + PC-MRI) when evaluating the performance of a BMHV.

**LIMITATIONS**

This study obviously includes limitations inherent to experimental studies on cardiovascular flows under healthy and pathological conditions. Although the results are based on time-resolved measurements at 400 Hz, they represent a 2D view of a more complex three-dimensional flow. The chosen plane is consistent with the one used in clinical imaging modalities like Doppler echocardiography and MRI. Future studies should however address this limitation by investigating the three-dimensional time-resolved flow structures using tomographic PIV. Our proposed approach based on the $\mu$-$\alpha$ plan has several advantages including: 1) it has the ability to differentiate between the dysfunction degree and the location of the dysfunctional leaflet; 2) it is angle independent and expected to be flow and valve size independent. However, one has to note from another side that performing such PC-MRI evaluations might be challenging and less accessible and accurate measurements require a good spatial resolution with an appropriate selection of the encoding velocity. Future studies have also to consider the time and financial aspects associated with the proposed approach. Some of the findings, more specifically the results related to wall shear stress and to the accumulation of viscous shear stresses, can also be challenging to be reproduced under *in vivo* settings since they require measurements with high temporal and spatial resolutions. Finally, in our study, we have tested six configurations including five dysfunctions. Although this appears to be enough to provide a good understanding of the flow characteristics in the ascending aorta in the presence of a dysfunctional BMHV, testing more configurations of valve leaflet dysfunctions, different valve sizes and designs will be useful in order to strengthen some of the findings (mostly the μ-α map and the correlation between mean values of α and the mean values of viscous energy dissipation).

**CONCLUSIONS**

In this study, the flow in the ascending aorta downstream of a dysfunctional bileaflet mechanical heart valve is investigated experimentally. Several configurations of leaflet dysfunctions are tested and the flow characteristics in the ascending aorta are evaluated in terms of viscous energy dissipation, wall shear stress and accumulation of viscous shear stresses. The results show that a bileaflet mechanical heart valve dysfunction leads to a complex spectrum of flow disturbances with each flow characteristic having its own worst case scenario in terms of dysfunction configuration. Furthermore, we introduce in this study a new approach based on the analysis of the normalized axial velocity profiles in the ascending aorta that has the potential to clearly discriminate not only between normal and dysfunctional bileaflet heart valves but also between the different leaflet dysfunction configurations. This approach could be easily implemented using phase-contrast MRI to follow up patients with bileaflet mechanical heart valves. Future *in vivo* studies are still required in order to confirm the findings of this experimental study.

**Table 1** Summary of working fluid properties and operating conditions.

| Working fluid properties | | Operating conditions | | Optical properties[#] | |
|---|---|---|---|---|---|
| Density ($\rho$)[*] | 1100 kg/m$^3$ | Cardiac Output | 4.3 ±0.12 L/min | RI (fluid) | 1.39 |
| Water- Glycerol Ratio | 60-40 (by volume) | Cardiac Cycle Period | 0.857 s | RI (acrylic) | 1.49 |
| Dynamic Viscosity[*] | 0.042 Pa.s | Heart Rate | 70 bpm | RI (mold) | 1.41 |
| | | Systolic Pressure | 125 ±3 mmHg | | |
| | | Diastolic Pressure | 70 ±3 mmHg | | |
| | | Average Systolic Duration | 0.376 s | | |
| | | Womersely Number (Wo) | 16.47 | | |

[*] Measured at 23°C.
[#] Refractive index at 23°C.

**Table 2** Summary of the different configurations of valve leaflet dysfunctions.

| Symbol | Leaflet dysfunction | | Leaflet opening angle (°) | |
|--------|---------------------|---|-------------|-----------------|
| | | | Sinus leaflet | Non-sinus leaflet |
| **N** | Normal | 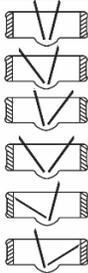 | 84.3 | 84.5 |
| **SLP** | Sinus Leaflet Partially blocked | 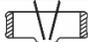 | 47.5 | 84.5 |
| **NSLP** | Non-Sinus Leaflet Partially blocked | 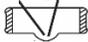 | 84.3 | 43 |
| **BLP** | Both Leaflets Partially blocked | 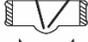 | 64.5 | 65 |
| **SLT** | Sinus Leaflet Totally blocked | 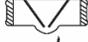 | 30[†] | 84.8 |
| **NSLT** | Non-Sinus Leaflet Totally blocked | 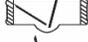 | 85 | 30[†] |

[†] Manufacturer value [46].

**Table 3** Flow characteritics computed from velocity field measurements

| Flow characteristics | |
| --- | --- |
| **Normalized axial velocity profile** | $V/V_{AVG} = \dfrac{V}{\frac{1}{2R}\int_{-R}^{R} V dr}$ |
| **Viscous energy dissipation** | $VED = \dfrac{1}{2}\rho \nu \sum_{i,j} \int \left(\dfrac{\delta u_i}{\delta x_j} + \dfrac{\delta u_j}{\delta x_i}\right)^2 dA$ |
| **Time-averaged wall shear stress** | $TAWSS = \dfrac{1}{T}\int_0^T |\bar{\tau}_w|\ dt$ |
| **Oscillatory shear index** | $OSI = \dfrac{1}{2}\left(1 - \dfrac{\left|\int_0^T \bar{\tau}_w\ dt\right|}{\int_0^T |\bar{\tau}_w|\ dt}\right)$ <br> $\bar{\tau}_w$ is the stress tensor |
| **Accumulation of viscous shear stress** | $\text{AVSS} = \sum \tau\ \Delta t$ <br> $\tau$ is the laminar viscous shear stress: <br> $\tau = \rho \nu \left(\dfrac{\delta u_i}{\delta x_j} + \dfrac{\delta u_j}{\delta x_i}\right)$ |

$\rho$ is the fluid density, $\nu$ is the kinematic viscosity, $u$ is the velocity in the $x$-direction, $v$ is the velocity in the $y$-direction, $\Delta t$ is the integration time step (20 μs), $T$ is the integration time (400 ms) and $dA$ is the area of the interrogation zone. AVSS is computed by advecting 1.8 10$^6$ particles in the flow field at the beginning of systole. TAWSS and OSI calculations where performed on filtered velocity fields (Savitzky-Golay for temporal noise filtering and proper orthogonal decomposition filtering for spatial noise). Spatial derivatives are computed using a compact-Richardson 4th order scheme [47]. More details can be found in Appendix A.

**Table 4** Summary of μ and α values for all investigated cases and the theoretical case.

| Case | $\mu$ | $\alpha$ |
|---|---|---|
| Theoretical | 0 | 0 |
| N | $\approx 0$ | $\approx 0$ |
| SLP | $> 0$ | $\gg 0$ |
| NSLP | $< 0$ | $\gg 0$ |
| BLP | $\approx 0$ | $\gg 0$ |
| SLT | $> 0$ | $\approx 1$ |
| NSLT | $< 0$ | $\approx 1$ |

**FIGURE LEGENDS**

**Figure 1** Experimental setup (A) shows the components of the double activated left heart duplicator with the LV being activated hydraulically by the linear motor driven piston while the LA is being activated passively by a servo motor driven cam follower arrangement. LV compliance is adjusted by controlling the air column height in the compliance chamber. (B) Left: camera alignment with the measurement plane, Center: details of the aortic model and its dimensions, Right: the six investigated cases: N; normal valve operation, SLP; sinus leaflet partially blocked, NSLP; non-sinus leaflet partially blocked, BLP; both leaflets partially blocked, SLT; sinus leaflet totally blocked, NSLT; non-sinus leaflet totally blocked.

**Figure 2** Normalized velocity profiles at the aortic root, sinotubular junction and downstream from the sinotubular junction for all cases during peak systole. N: Normal; SLP: Sinus Leaflet Partially blocked; NSLP: Non-Sinus Leaflet Partially blocked; BLP: Both Leaflets Partially blocked; SLT: Sinus Leaflet Totally blocked; NSLT: Non-Sinus Leaflet Totally blocked.

**Figure 3** Temporal evolution of space-averaged viscous energy dissipation per unit of depth for all the cases. The insert shows the systolic average for each case. N: Normal; SLP: Sinus Leaflet Partially blocked; NSLP: Non-Sinus Leaflet Partially blocked; BLP: Both Leaflets Partially blocked; SLT: Sinus Leaflet Totally blocked; NSLT: Non-Sinus Leaflet Totally blocked.

**Figure 4** Time-averaged wall shear stress (solid line) and oscillatory shear index (dashed line) on the sinus wall (left column) and on the non-sinus wall (right column) for all cases investigated. N: Normal; SLP: Sinus Leaflet Partially blocked; NSLP: Non-Sinus Leaflet Partially blocked; BLP: Both Leaflets Partially blocked; SLT: Sinus Leaflet Totally blocked; NSLT: Non-Sinus Leaflet Totally blocked.  The dash dot line refers to OSI =0.5.

**Figure 5** Histograms of the accumulation of viscous shear stresses on advected tracers released at the beginning of systole. N: Normal; SLP: Sinus Leaflet Partially blocked; NSLP: Non-Sinus Leaflet Partially blocked; BLP: Both Leaflets Partially blocked; SLT: Sinus Leaflet Totally blocked; NSLT: Non-Sinus Leaflet Totally blocked.

**Figure 6** Skewness mean values (blue bar) and standard deviations (orange lines) for all cases. N: Normal; SLP: Sinus Leaflet Partially blocked; NSLP: Non-Sinus Leaflet Partially blocked; BLP: Both Leaflets Partially blocked; SLT: Sinus Leaflet Totally blocked; NSLT: Non-Sinus Leaflet Totally blocked.

**Figure 7** Alpha-skewness map for all cases. The map shows the mean values of alpha and skewness for each case (marked with the circle) while the standard deviation for alpha and skewness is shown vertically and horizontally respectively. N: Normal; SLP: Sinus Leaflet Partially blocked; NSLP: Non-Sinus Leaflet Partially blocked; BLP: Both Leaflets Partially blocked; SLT: Sinus Leaflet Totally blocked; NSLT: Non-Sinus Leaflet Totally blocked.

**Figure 8** Correlation between mean values of $\alpha$ and mean values of viscous energy dissipation (R=0.92).

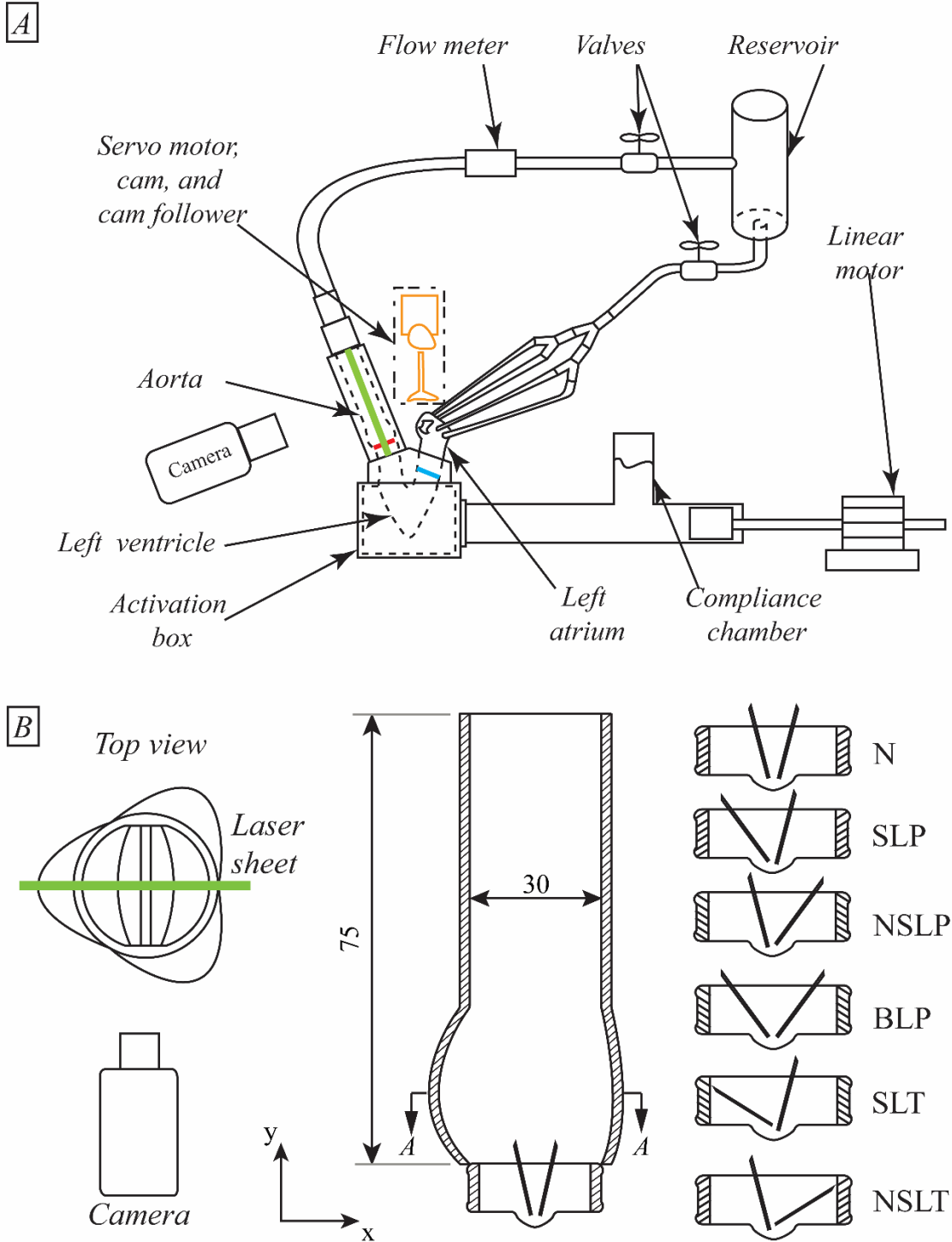

**Figure 1** Experimental setup (A) shows the components of the double activated left heart duplicator with the LV being activated hydraulically by the linear motor driven piston while the LA is being activated passively by a servo motor driven cam follower arrangement. LV compliance is adjusted by controlling the air column height in the compliance chamber. (B) Left: camera alignment with the measurement plane, Center: details of the aortic model and its dimensions, Right: the six investigated cases: N; normal valve operation, SLP; sinus leaflet partially blocked, NSLP;

non-sinus leaflet partially blocked, BLP; both leaflets partially blocked, SLT; sinus leaflet totally blocked, NSLT; non-sinus leaflet totally blocked.

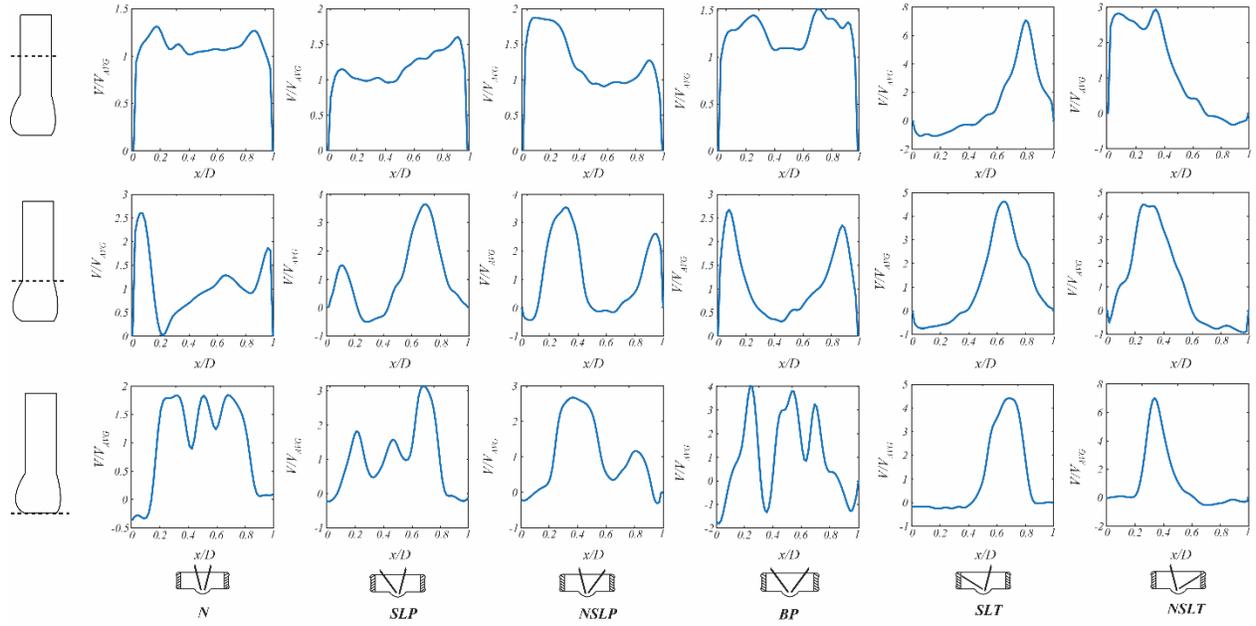

**Figure 2** Normalized velocity profiles at the aortic root, sinotubular junction and downstream from the sinotubular junction for all cases during peak systole. N: Normal; SLP: Sinus Leaflet Partially blocked; NSLP: Non-Sinus Leaflet Partially blocked; BLP: Both Leaflets Partially blocked; SLT: Sinus Leaflet Totally blocked; NSLT: Non-Sinus Leaflet Totally blocked.

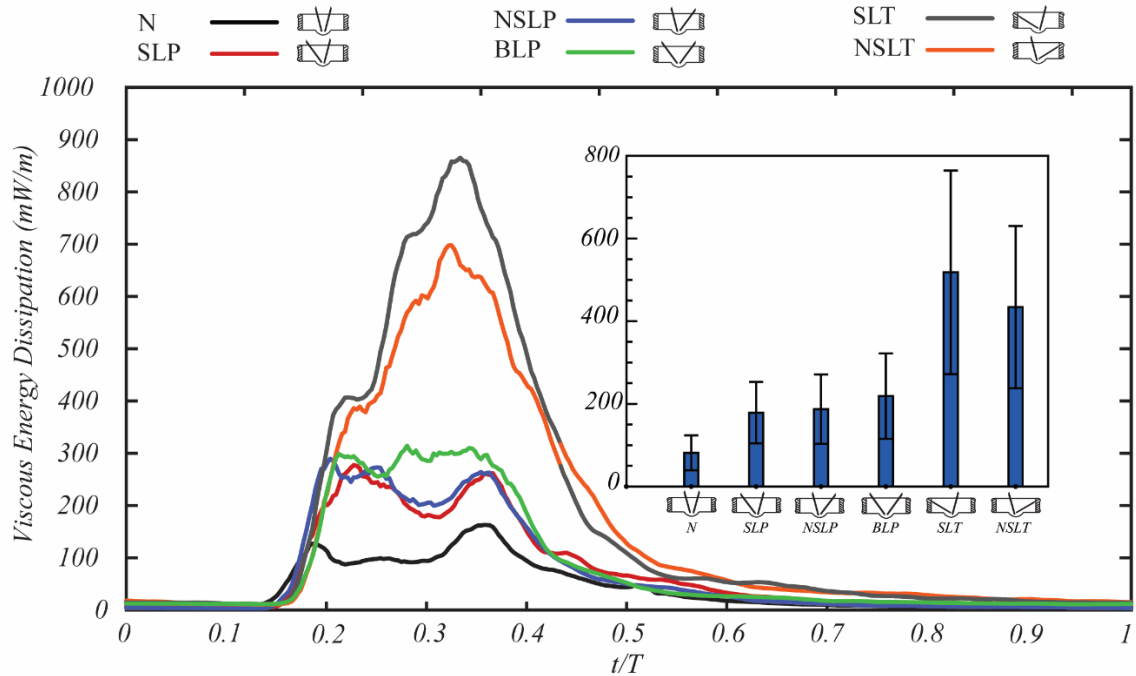

**Figure 3** Temporal evolution of space-averaged viscous energy dissipation per unit of depth for all the cases. The insert shows the systolic average for each case. N: Normal; SLP: Sinus Leaflet Partially blocked; NSLP: Non-Sinus Leaflet Partially blocked; BLP: Both Leaflets Partially blocked; SLT: Sinus Leaflet Totally blocked; NSLT: Non-Sinus Leaflet Totally blocked.

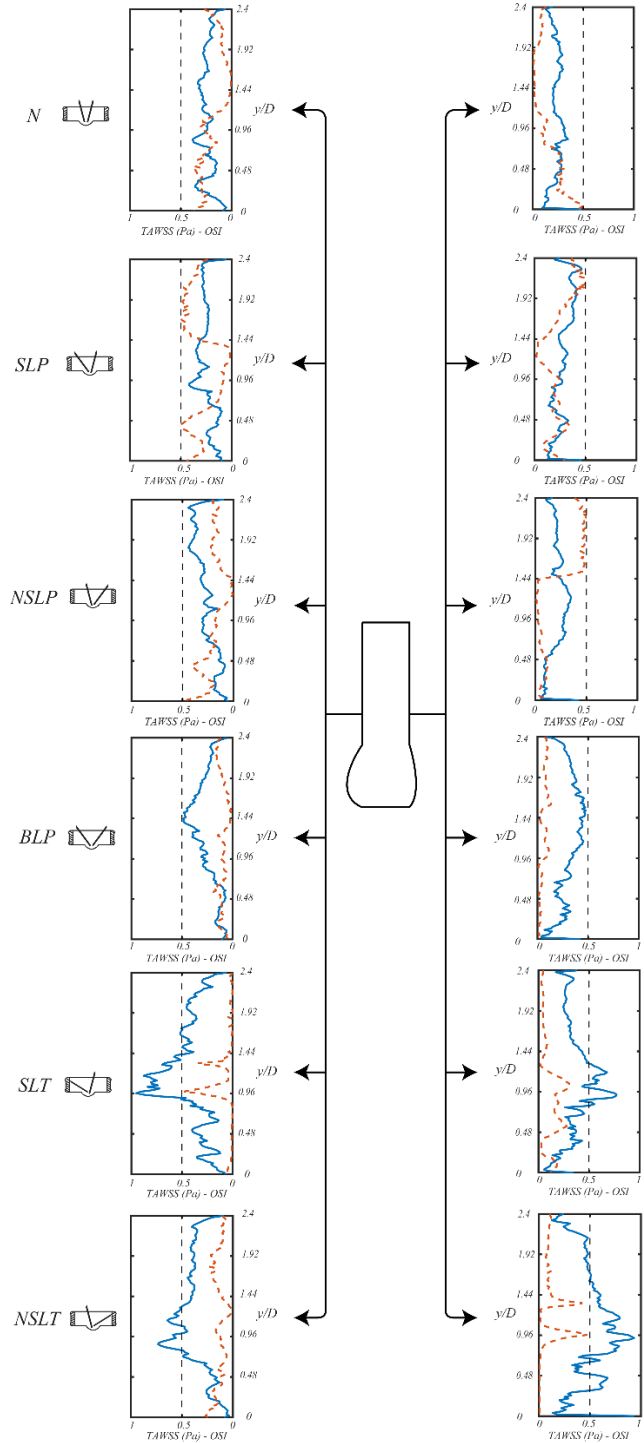

**Figure 4** Time-averaged wall shear stress (solid line) and oscillatory shear index (dashed line) on the sinus wall (left column) and on the non-sinus wall (right column) for all cases investigated. N: Normal; SLP: Sinus Leaflet Partially blocked; NSLP: Non-Sinus Leaflet Partially blocked; BLP: Both Leaflets Partially blocked; SLT: Sinus Leaflet Totally blocked; NSLT: Non-Sinus Leaflet Totally blocked. The dash dot line refers to OSI =0.5.

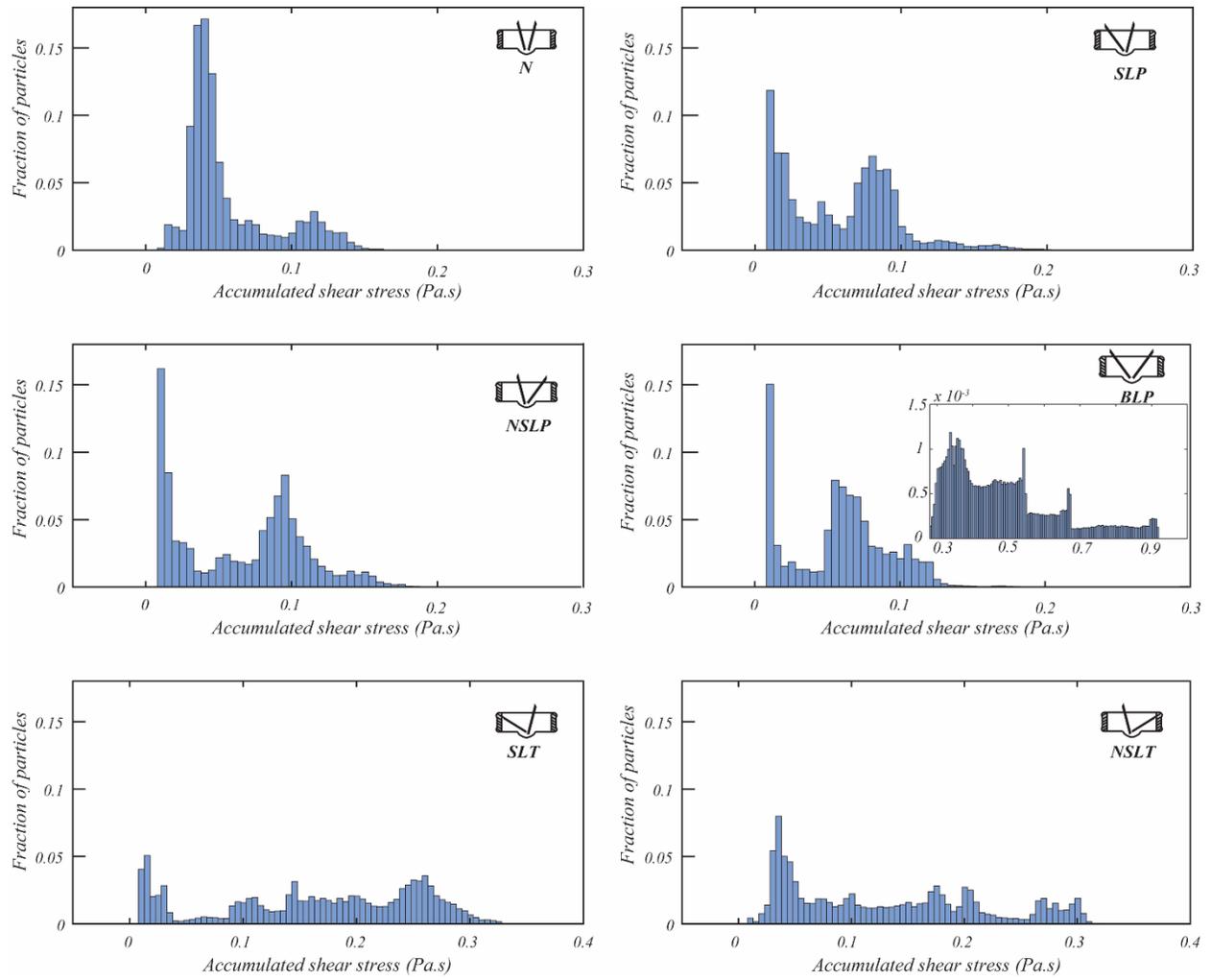

**Figure 5** Histograms of the accumulation of viscous shear stresses on advected tracers released at the beginning of systole. N: Normal; SLP: Sinus Leaflet Partially blocked; NSLP: Non-Sinus Leaflet Partially blocked; BLP: Both Leaflets Partially blocked; SLT: Sinus Leaflet Totally blocked; NSLT: Non-Sinus Leaflet Totally blocked.

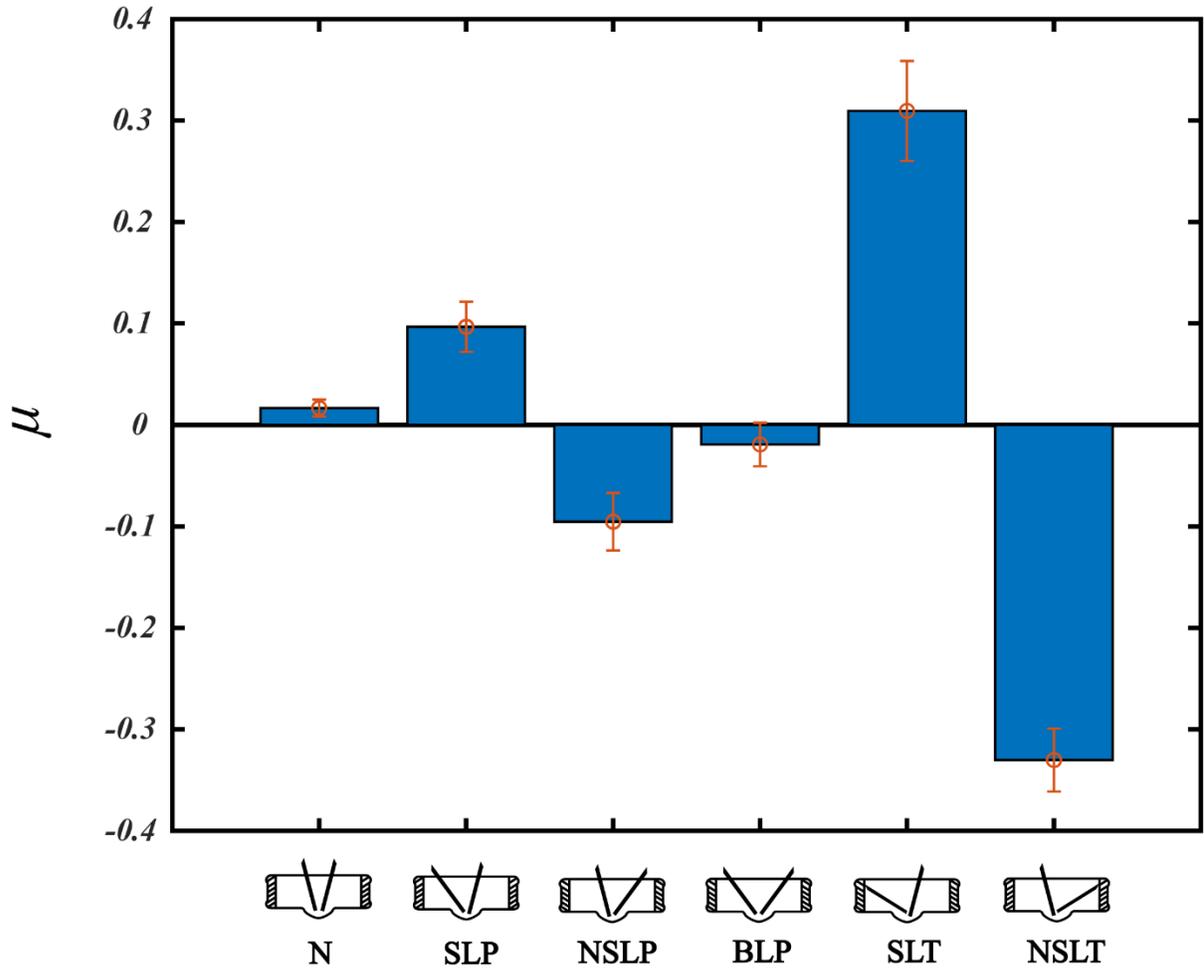

**Figure 6** Skewness mean values (blue bar) and standard deviations (orange lines) for all cases. N: Normal; SLP: Sinus Leaflet Partially blocked; NSLP: Non-Sinus Leaflet Partially blocked; BLP: Both Leaflets Partially blocked; SLT: Sinus Leaflet Totally blocked; NSLT: Non-Sinus Leaflet Totally blocked.

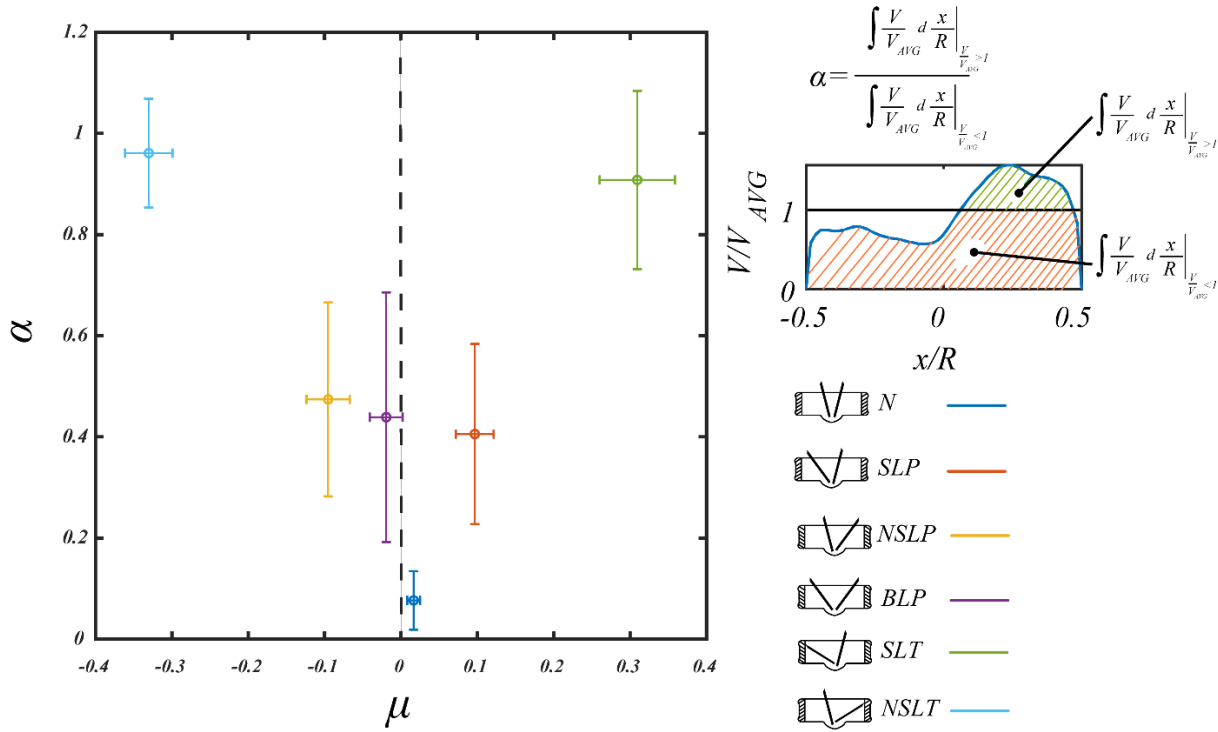

**Figure 7** Alpha-skewness map for all cases. The map shows the mean values of alpha and skewness for each case (marked with the circle) while the standard deviation for alpha and skewness is shown vertically and horizontally respectively. N; normal valve operation, SLP; sinus leaflet being partially restricted, NSLP; non-sinus leaflet being partially restricted, BLP; both leaflets being partially restricted, SLT; sinus leaflet being totally restricted, NSLT; non-sinus leaflet being totally restricted.

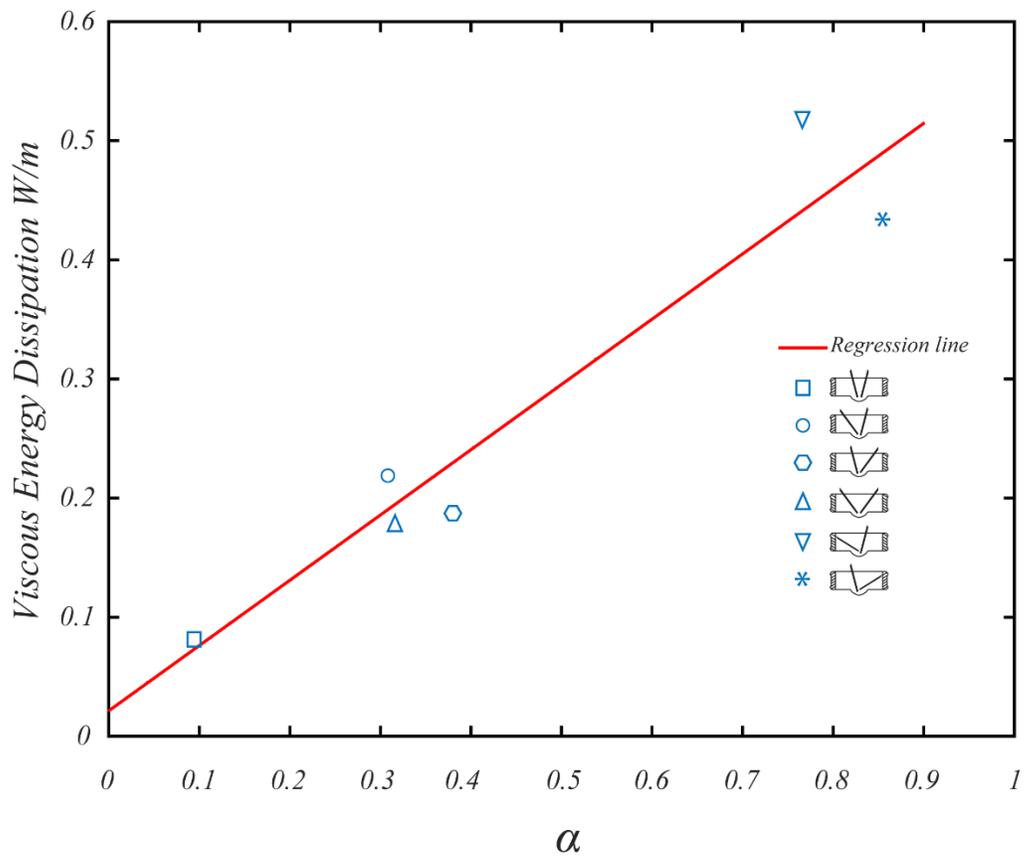

**Figure 8** Correlation between mean values of α and mean values of viscous energy dissipation (R=0.92).

# APPENDIX A

***Elastic models.*** In this study, a model of an aorta including the sinuses of Valsalva is constructed using silicone molding. Up to four layers of silicone (XIAMETER RTV-4234-T4) are coated on a 3D-printed core mold of the aorta. The final thickness of the silicone aorta is 2±0.5 mm. The cured silicone has a Young modulus of elasticity of 1.675 MPa and a refractive index of 1.41. The same procedure was used in order to create elastic models of the left ventricle and the left atrium.

***In vitro heart simulator.*** In our setup, the model of the left ventricle is enclosed in a rigid Plexiglas box connected to a piston-cylinder assembly. The piston cylinder assembly is driven by a linear motor (Servo Drive E1100-RS, NTI AG; Switzerland). The controlled movement of the piston generates a cyclic contraction and expansion of the elastic model of the left ventricle simulating cardiac systolic and diastolic phases. The working fluid is a mixture of water/glycerol with a volumetric ratio of 60% / 40% respectively. The flow rate in the model is recorded using a magnetic inductive flow sensor (ProSense FMM50-102, Germany, accuracy ±0.12 L/min) while the pressure is recorded using a fiber optic pressure sensor (FISO FOP-M260, Canada; range -300 to 300 mmHg; resolution < 3 mmHg.

***Viscous energy dissipation (VED).*** Blood flowing through a dysfunctional  BMHV is expected to lead to viscous energy losses where large unstable eddies are formed (when the flow hits the dysfunctional leaflets) and will form smaller eddies that will keep dissipating the flow energy and convert it to heat [1]. This form of energy is unrecoverable [2]. For each case tested in this study, space-averaged VED are calculated using the time-resolved velocity fields obtained in the ascending aorta: $VED = \frac{1}{2} \rho \nu \sum_{i,j} \int \left( \frac{\delta u_i}{\delta x_j} + \frac{\delta u_j}{\delta x_i} \right)^2 dA$ . Since there exists a linear relationship between transvalvular pressure gradients and VED as reported by [3,4], then, the energy dissipation has to be compensated by the LV in order maintain a normal heart pumping function.

***Time-averaged wall shear stress (TAWSS) and oscillatory shear index (OSI).*** Several studies have reported the direct link between abnormal flow patterns in the ascending aorta, mostly eccentric flow jets, and the development of aneurysms of the ascending aorta [5–7]. Most of those studies focused on bicuspid aortic valves. In the presence of a BMHV dysfunction, it is anticipated that the flow field in the ascending aorta will experience significant changes compared to a heathy native flow field. This will subject the aortic wall to non-physiological loadings. In our study, the TAWSS and OSI values were obtained by post-processing time-resolved velocity fields. It is important to note at this stage that pre-processing steps are needed in order to reduce spatial-noise and temporal-noise in the instantaneous velocity fields. Proper orthogonal decomposition (POD) was used as a filtering method in order to reduce the spatial-noise in the instantaneous velocity fields as suggested by [8]. The temporal-noise in the instantaneous velocity fields was reduced first by applying a Savitzky-Golay filter [9] followed by POD filtering of spatial-noise. The reconstructed velocity fields are then interpolated on a more refined grid using two-dimensional cubic-spline interpolation. The refined grid has a spatial resolution of 55 $\mu$m. Second order forward and backward finite differences are finally applied on both walls to compute the velocity gradient near the wall. Since the wall is moving, wall detection is performed using a custom-made Matlab code where the wall is identified based on the difference in intensities across the wall at each captured frame. The angle ($\theta$) between the x-y coordinate system and the normal direction on each point on the wall is evaluated. Then, the shear stress calculated using $\tau_{ij} = \rho \nu \left( \frac{\delta u_i}{\delta x_j} + \frac{\delta u_j}{\delta x_i} \right)$ is subjected to a rotational transformation based on $\theta$ yielding the rotated shear stress tensor $\bar{\tau}_{ij}$:

$$\bar{\tau}_{ij} = \begin{bmatrix} \cos\theta & \sin\theta \\ -\sin\theta & \cos\theta \end{bmatrix} \tau_{ij} \begin{bmatrix} \cos\theta & -\sin\theta \\ \sin\theta & \cos\theta \end{bmatrix}$$

By considering the tangential components of the stress tensor $\bar{\tau}_w$, time-averaged wall shear stress (TAWSS) and oscillating shear index (OSI) are then calculated for each point according to:

$$TAWSS = \frac{1}{T}\int_0^T |\bar{\tau}_w| \ dt$$

$$OSI = \frac{1}{2}\left(1 - \frac{\left|\int_0^T \bar{\tau}_w \ dt\right|}{\int_0^T |\bar{\tau}_w| \ dt}\right)$$

***Accumulation of viscous shear stresses (AVSS).*** Platelet activation is known to be the major stimuli of thrombus formation. According to Hellums et al. [10], platelet activation is related to the applied shear stress and exposure time. Ge et al. [11] also showed that viscous shear stresses are the major stresses applied on platelets while Reynolds shear stresses are simply to be considered as a turbulence statistical tool. In our study, AVSS is computed using the recorded time-resolved velocity fields according to Bluestein et al. model [12]. A rectangular grid of equally spaced tracers, placed directly above the BMHV, is released within the time resolved velocity fields. Each point is following a pathline which is computed from its velocity gradient tensor. Time-stepping is performed using fourth-order Runge-Kutta scheme, while the particle inertia is ignored. The history of the tracer's position is then used to extract its viscous shear stress ($\tau$) value at each instant. The computed AVSS could represent, therefore, an approximation of the viscous shear stress environment surrounding platelets. The advection is performed at the beginning of the systolic phase and more than $1.8 \times 10^6$ tracers are released. The initial spacing between tracers is 20 μm in $x$ and $y$ directions. The integration time for particle advection is 400 ms with a time interval of 2.5 µs. The AVSS is computed for the tracers that remained for the whole advection duration in the region of interest.

# APPENDIX B

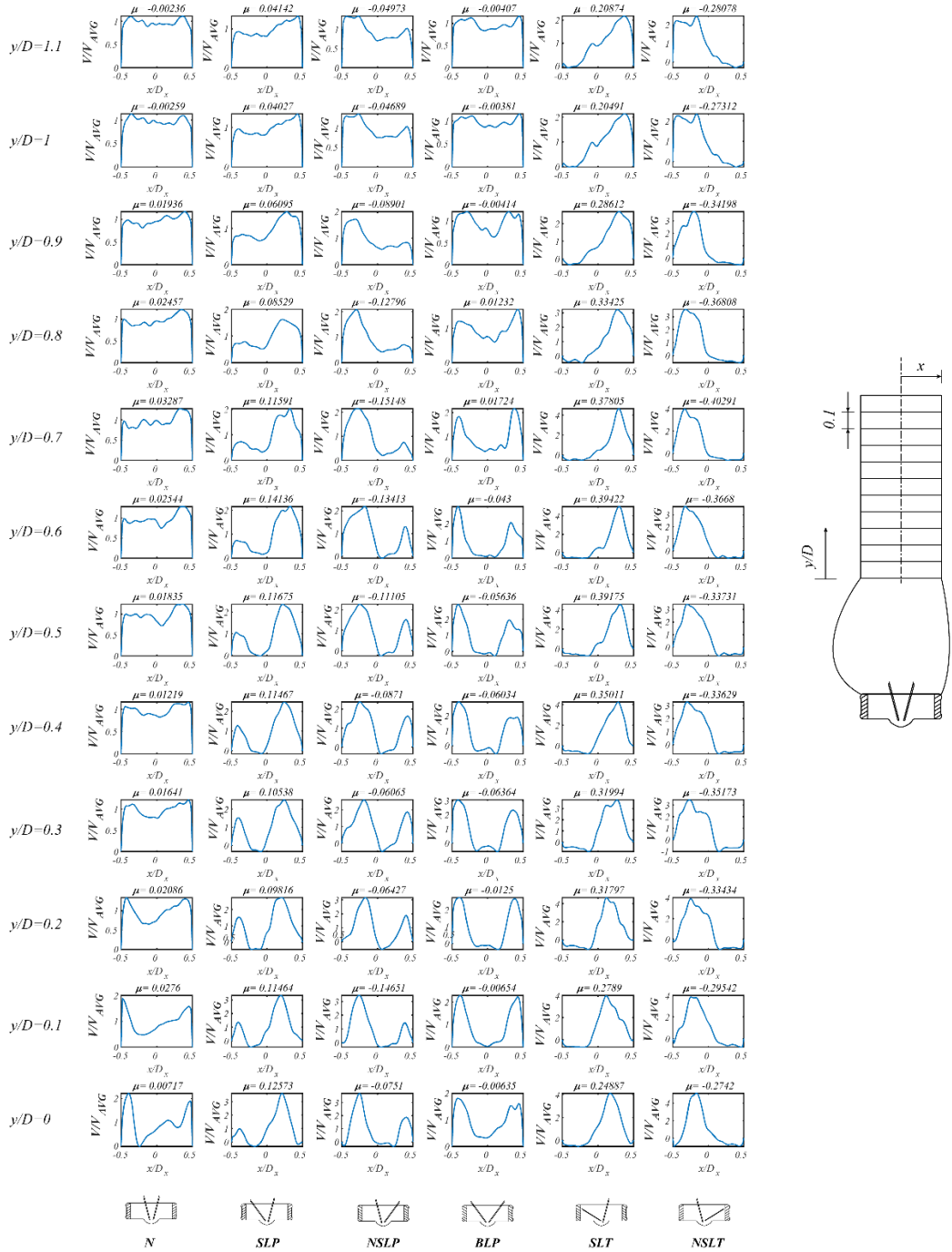

**Figure B.1** Normalized axial velocity profiles at twelve axial positions at the peak of systole starting. The axial velocity profile is normalized by the average axial velocity for each case, while μ refers to the profile skewness. N: Normal; SLP: Sinus Leaflet Partially blocked; NSLP: Non-Sinus Leaflet Partially blocked; BLP: Both Leaflets Partially blocked; SLT: Sinus Leaflet Totally blocked; NSLT: Non-Sinus Leaflet Totally blocked.